\documentclass[footinbib,twocolumn,amsmath,amstex,amssymb,mathfonts,superscriptaddress,prl]{revtex4}
\usepackage{graphicx}
\usepackage{color}
\usepackage{bm}
\usepackage{verbatim}
\usepackage{ulem}
\usepackage{marvosym}

\usepackage{amsmath}
\usepackage{amssymb}
\usepackage{amsthm}
\usepackage{amsfonts}
\usepackage[caption=false]{subfig}

\newcommand{\be}{\begin{equation}}
\newcommand{\ee}{\end{equation}}
\newcommand{\bea}{\begin{eqnarray}}
\newcommand{\eea}{\end{eqnarray}}
\def\nablabold{\mbox{\boldmath $\nabla$}}

\newcommand{\Mv}{{\bf M}}

\newcommand{\jv}{{\bf j}}
\newcommand{\rv}{{\bf r}}

\newcommand{\kv}{{\bf k}}
\newcommand{\zv}{{\bf z}}

\renewcommand{\vec}[1]{{\bf #1}}
\def\nn{\nonumber\\}

\def\bef{\begin{framed}}
\def\eef{\end{framed}}
\def\be{\begin{equation}}
\def\ee{\end{equation}}
\def\ber{\begin{eqnarray}}
\def\eer{\end{eqnarray}}

\def\nablabold{\mbox{\boldmath $\nabla$}}

\def\rv{{\bf r}}

\def\Mv{{\bf M}}

\def\jv{{\bf j}}

\def\kv{{\bf k}}

\def\nn{\nonumber}

\begin{document}
\title{Hall diffusion anomaly and transverse Einstein relation} 
\date{\today}
\author{Justin C. W. Song} 
\email{justinsong@ntu.edu.sg}
\affiliation{Division of Physics and Applied Physics, Nanyang Technological University, Singapore 637371}
\affiliation{Institute of High Performance Computing, Agency for Science, Technology, \& Research, Singapore 138632}
\author{Giovanni Vignale}
\email{vignaleg@missouri.edu}
\affiliation{Department of Physics and Astronomy, University of Missouri, Columbia, Missouri 65211,~USA}
\affiliation{Center for Advanced 2D materials, National University of Singapore, Singapore 117542}

\begin{abstract}
It is commonly believed that the current response of an electron fluid to a mechanical force (such as an electric field) or to a ``statistical force" (e.g., a gradient of chemical potential) are governed by a single linear transport coefficient - the electric conductivity.  We argue that this is not the case in anomalous Hall materials.
In particular, we find that transverse (Hall) currents manifest two distinct Hall responses governed by an unconventional {\it transverse} Einstein relation that captures an anomalous relation between the Hall conductivity and the Hall diffusion constant. 
We give examples of when the Hall diffusion anomaly is prominent, resulting in situations where the transverse diffusion process overwhelms the Hall conductivity and vice versa. 
\end{abstract} 
\maketitle

Two types of forces can drive currents in electron fluids: mechanical and statistical. A familiar example of the first kind is the force exerted by an electric field, giving rise to  normal electrical conduction in a metal. Such a force is typically described by a potential term in the Hamiltonian. An example of the second kind is the force generated by a gradient of chemical potential, giving rise to a diffusion current according to Ficks' law\cite{ashcroftmermin}.  In this case the hamiltonian is unchanged, but the system spontaneously evolves towards the equilibrium state starting from initial non-equilibrium conditions.

It is a tenet of transport theory that the responses to mechanical and statistical forces (under standard assumptions of linear response and slow spatial variations) are identical~\cite{ashcroftmermin, girvinkunyang, kubo}. Here we show that this is not the case: a difference between the responses to mechanical and statistical forces arises when one considers the {\it transverse} response (transverse relative to the direction of the applied force) in materials whose band structure exhibits Berry curvature, such as anomalous Hall systems~\cite{Ong,dixiao10} and other topological materials~\cite{dixiao07,gorbachev14,tarucha15,zhang15,sid}. We term this a ``Hall diffusion anomaly''. 

We arrive at the Hall diffusion anomaly using general equilibrium and macroscopic considerations. Indeed, the ensuing relationship between responses to mechanical and statistical forces in anomalous Hall systems is universal, i.e., independent of microscopic details of the hamiltonian (such as the presence or absence of disorder) and furthermore it is applicable to {\it non-equilibrium} as well as equilibrium conditions. Strikingly, we find that when the Hall diffusion anomaly is pronounced, materials can exhibit transverse diffusion which overwhelms conventional Hall conductivity (response to a mechanical force); conversely, we find other cases wherein conventional Hall conductivity is large but transverse diffusion is arrested. We anticipate the implications will be even larger when dealing with thermal transport, or the transport of charge neutral particles, where, arguably, only the statistical force (the gradient of the temperature) exists.

{\bf Drift and Diffusion currents.} When a spatially varying electric potential $\delta\varphi(\rv)$ and chemical potential $\delta \mu(\vec r)$ are applied to a metal, the induced charge current density at any given position $\vec r$ is given  by the local relation
\be
\delta \jv(\rv) = \widehat{\boldsymbol{\mathcal{L}}} \big[-\nablabold\delta\varphi(\rv)\big] + \widehat{\boldsymbol{\mathcal{N}}}\big[\nablabold \delta \mu(\vec r)/e \big], 
\label{eq:fullcurrent}
\ee
where the tensors $\widehat{\boldsymbol{\mathcal{L}}}$ and $\widehat{\boldsymbol{\mathcal{N}}}$ describe the (linear response) current density that develops, and $-e$ is the electron charge. We have kept only the lowest order symmetry-allowed terms and assumed slow spatial variations.  

While the tensors $\widehat{\boldsymbol{\mathcal{L}}}$ and $\widehat{\boldsymbol{\mathcal{N}}}$ can, a priori, be different, a strong local relation between the two can be established. This arises from the observation that simultaneous variations of $\varphi (\vec r)$ and $\mu (\vec r)$,  which leave the electrochemical potential $-e \phi  (\vec r)+ \mu (\vec r)$ unchanged, do not change the {\it longitudinal} component of the current~\cite{ashcroftmermin, girvinkunyang, kubo}. This yields 
\be\label{EinsteinRelationLongitudinal}
\widehat{\boldsymbol{\mathcal{L}}}_L=\widehat{\boldsymbol{\mathcal{N}}}_L\,,
\ee
where the subscript $L$ specifies the longitudinal component of the corresponding tensor. This is the familiar Einstein relation, connecting the longitudinal conductivity to the longitudinal diffusion constant~\cite{ashcroftmermin, girvinkunyang, kubo}.  

Our main result in this paper is that the chain of reasoning leading to Eq.~(\ref{EinsteinRelationLongitudinal})  fails for the {\it transverse} components of the response tensors, yielding significant and observable differences between the transverse responses to mechanical and statistical forces, i.e., 
\be
\widehat{\boldsymbol{\mathcal{L}}}_H\neq\widehat{\boldsymbol{\mathcal{N}}}_H\,,
\label{eq:EinsteinViolation}
\ee
where the subscript $H$ (from Hall) denotes the transverse components. 

As we explain below, this violation -- the Hall diffusion anomaly -- naturally manifests in systems with broken time-reversal symmetry, such as anomalous Hall materials~\cite{Ong,dixiao10} carrying a spontaneous magnetization, and also, more subtly, in time-reversal invariant multi-valley/flavor systems~\cite{dixiao07,gorbachev14,tarucha15,zhang15,sid}.  
In all these systems, we show that the difference between 
$\widehat{\boldsymbol{\mathcal{L}}}_H$ and $\widehat{\boldsymbol{\mathcal{N}}}_H$  leads to a novel formula for the anomalous Hall diffusion constant (or the transverse valley/flavor diffusion constant), which is quantitatively and qualitatively different from the formula that would be obtained by a 
direct extrapolation of the longitudinal Einstein relation. 

{\bf Hall diffusion anomaly.} For the sake of clarity and simplicity we present our derivation for two-dimensional anomalous Hall systems, and fix the magnetization $\Mv (\vec r)$ perpendicular to the plane: $\Mv (\vec r) =M (\vec r) \hat \zv$, $\hat\zv$ is the unit vector perpendicular to the plane; throughout we focus only on orbital magnetization with spin a spectator degree of freedom. A formulation in three dimensions is presented in the Supplementary Information, {\bf SI}~\cite{SI}. 

Assuming local isotropy, we write the tensor $\widehat{\boldsymbol{\mathcal{L}}}$ as 
\be\label{DriftCurrent1}
\widehat{\boldsymbol{\mathcal{L}}} \big[-\nablabold\delta\varphi(\rv)\big]  = -\sigma_L\nablabold\delta\varphi(\rv)-\sigma_H \hat{\zv}\times\nablabold\delta\varphi(\rv)\,.  
\ee
with longitudinal conductivity $\sigma_L$  and  transverse (Hall) conductivity $\sigma_H$. In a similar fashion, we express the tensor $\widehat{\boldsymbol{\mathcal{N}}}$ in terms of a longitudinal diffusion constant $D_L$ and a transverse (Hall) diffusion constant $D_H$: 
\begin{eqnarray}
\label{DiffusionCurrent1}
\widehat{\boldsymbol{\mathcal{N}}}\big[\nablabold \delta \mu(\vec r) \big] =e\frac{\partial n}{\partial\mu}\left[D_L \nablabold \delta \mu (\vec r) + D_H\hat{\zv}\times \nablabold \delta \mu (\vec r) \right], 
\end{eqnarray}
where the thermodynamic derivative of the electron density with respect to chemical potential $\partial n/\partial \mu$ is taken in the uniform state at the local equilibrium density. Recalling $(\partial n/\partial \mu)\nablabold\mu(\rv)=\nablabold \delta n(\rv)$, it is clear that Eq.~(\ref{DiffusionCurrent1}) is indeed a diffusion current proportional to the gradient of the density.

Crucially, Eqs.~(\ref{eq:fullcurrent}),~(\ref{DriftCurrent1}), and~(\ref{DiffusionCurrent1}) give the (linear response) current density for arbitrary spatial distributions of $\delta \varphi(\vec r)$ and $\delta \mu(\vec r)$. As such, these apply for out-of-equilibrium situations as well as in circumstances when the electron gas is allowed to relax back into equilibrium. Indeed, as long as the system is in the regime of linear response, both out-of-equilibrium and in-equilibrium circumstances [which can possess very different $\varphi(\vec r)$ and $\mu(\vec r)$] possess the {\it same} coefficients $\widehat{\boldsymbol{\mathcal{L}}}$ and $\widehat{\boldsymbol{\mathcal{N}}}$~\cite{ashcroftmermin,girvinkunyang}. As we now argue, this constancy of $\widehat{\boldsymbol{\mathcal{L}}}$ and $\widehat{\boldsymbol{\mathcal{N}}}$ in the linear response regime allows us to derive relations between the two transport coefficients. In what follows, our strategy is to consider the profile of current density at equilibrium where there are stringent thermodynamic constraints that relate the form of current density, the spatially varying carrier density, and the electric potential.

To proceed, we apply a spatially varying electric potential $\delta \varphi(\rv)$ 
{\it and} allow the electron gas to relax to equilibrium. In such an equilibrium situation, the electro-chemical potential is uniform across all space~\cite{ashcroftmermin,girvinkunyang}, i.e., 
\be\label{EquilibriumCondition}
e\nablabold \delta \varphi(\rv) = \nablabold \delta \mu^{\rm eq}(\rv)\,,
\ee
where the superscript $``{\rm eq}"$ denotes the chemical potential profile that the electron gas adopts when it is allowed to relax to equilibrium.  
Under such variations, Eq.~(\ref{EquilibriumCondition}), the system remains in equilibrium, 
and the longitudinal component of the charge current density must therefore vanish, see subscript ``$L$'' terms in Eq.~(\ref{DriftCurrent1}) and~(\ref{DiffusionCurrent1}).

In contrast, the transverse component of the current density does not necessarily vanish, and changes 
as $\delta \varphi(\vec r)$ and $\delta \mu^{\rm eq} (\vec r)$ profiles [satisfying Eq.~(\ref{EquilibriumCondition})] are varied.
This is exactly what happens in anomalous Hall systems, where the induced change in equilibrium current density is 
\be\label{deltaj-equilibrium}
\delta \jv_{\rm eq}(\rv)=\nablabold \times \left[\delta M_{\rm eq}(\rv) \hat{\vec{z}}\right] = \nablabold \delta M_{\rm eq}(\rv) \times \hat{\vec{z}}\,,
\ee
where $\delta M_{eq}(\rv)$ is the change in equilibrium magnetization (between distinct equilibrium states) induced by the applied 
$\delta \varphi(\vec r)$. 

The gradient of the equilibrium magnetization is related to the gradient of the chemical potential as follows:
 \be
 \label{eq:dmdmu}
\nablabold \delta M_{\rm eq}(\rv)= \left(\frac{\partial M}{\partial \mu}\right)_B \nablabold \delta \mu^{\rm eq}(\rv) = \left(\frac{\partial n}{\partial B}\right)_\mu \nablabold \delta \mu^{\rm eq}(\rv)\,.
\ee
In the first equality, the derivative of $M$ with respect to a local equilibrium $\mu$ is calculated in thermodynamic equilibrium at constant magnetic field $B$.  The second equality follows from a Maxwell relation, and the derivative of the equilibrium density with respect to magnetic field is calculated at constant chemical potential~\cite{Kardar,streda2}.   

We now require that the sum of the linear responses, 
Eqs. ~(\ref{DriftCurrent1}) and (\ref{DiffusionCurrent1}), equal the $\delta \varphi(\vec r)$-induced change in the equilibrium current density, Eq.~(\ref{deltaj-equilibrium}), 
when the equilibrium condition~(\ref{EquilibriumCondition}) is satisfied. This leads us directly to the relations
\be\label{EinsteinL}
\sigma_L=e^2 \frac{\partial n}{\partial\mu} D_L \,,
\ee
\be\label{EinsteinH}
\sigma_H-e\left(\frac{\partial n}{\partial B}\right)_\mu = e^2\frac{\partial n}{\partial\mu} D_H \,.
\ee
The first relation, Eq.~(\ref{EinsteinL}), is the usual Einstein relation for the longitudinal diffusion constant. 

The second relation, Eq.~(\ref{EinsteinH}), is the transverse Einstein relation for the anomalous Hall conductivity and the anomalous Hall diffusion constant in systems in which the equilibrium current does not vanish. It encapsulates the Hall diffusion anomaly -- Eq.~(\ref{eq:EinsteinViolation}) -- where mechanical and statistical forces give contrasting current responses; it violates the simple expectations that the Hall diffusion and Hall conductivity are directly proportional to each other (e.g., if Eq.~(\ref{EinsteinL}) were directly applied to the transverse components). 

Key to this anomaly is $e(\partial n/\partial B)_\mu$ in Eq.~(\ref{EinsteinH}): this captures the changes in the equilibrium current (via equilibrium magnetization) as chemical potential is varied [see Eq.~(\ref{eq:dmdmu})]. Since the linear responses 
Eqs.~(\ref{DriftCurrent1}) and (\ref{DiffusionCurrent1}) describe the current response even under equilibrium conditions, the only way in which the equilibrium magnetization can change as chemical potential is varied is from an imbalance in drift and diffusion currents~(e.g. as shown for massive Dirac fermions~\cite{Pesin}). This explains the origin of the Hall diffusion anomaly. Interestingly, when the density of states $\partial n/\partial\mu$ vanishes in a bulk spectral gap,  Eq.~(\ref{EinsteinH}) reproduces the familiar Str\u eda formula for the quantized Hall conductivity of a bulk incompressible system~\cite{streda2}. Indeed, if $e(\partial n/\partial B)_\mu$ were absent, a na\"ive application of Eq.~(\ref{EinsteinL}) to the Hall conductivity would lead to the paradoxical result that $D_H=\infty$ for a quantum anomalous Hall insulator. 

Two important comments are now in order. The first is that the relationship between $\sigma_H$ and $\bar \sigma_H$ (or $D_H$), while deduced from equilibrium considerations, is valid for general {\it non-equilibrium} situations. This is because $\widehat{\boldsymbol{\mathcal{L}}}$  and $\widehat{\boldsymbol{\mathcal{N}}}$ in Eq.~(\ref{eq:fullcurrent}) remain the same in the regime of linear response -- they do not change between equilibrium or non-equilibrium situations. Therefore, our distinction between Hall currents (driven by an electric field) and Hall diffusion currents (driven by a gradient of density) goes well beyond the previously recognized distinction between {\it equilibrium currents} flowing in the incompressible and compressible regions of a quantum Hall channel~\cite{Geller94,Uri2020}.

Second, the relation Eq.~(\ref{EinsteinH}) does not depend on the microscopic details of the Hamiltonian. For example, it is known that the anomalous Hall conductivity of massive Dirac fermions (an 
example discussed below) is affected by disorder~\cite{Sinitsyn2007, Ong, Titov2015}. Yet because Eq.~(\ref{EinsteinH}) arises under general macroscopic drift-diffusion considerations, it remains valid and applicable even in disordered systems. Indeed, Eq.~(\ref{EinsteinH}) can be readily employed to relate the Hall conductivity and diffusion constants provided the same model is consistently used for the microscopic calculation of $\sigma_H$,$\bar \sigma_H$, and $(\partial n/\partial B)_\mu$. 

{\bf Fermi surface magnetic moment and Hall diffusion.} The difference between the Hall drift current (mechanical response) and the Hall diffusion current (statistical response) captured by Eq.~(\ref{EinsteinH}) is further underscored by the distinct origins of $\sigma_H$ and $D_H$. To demonstrate this, 
we concentrate on crystalline anomalous Hall systems with electrons hosted in bulk Bloch bands. The equilibrium density as a function of chemical potential and magnetic field in the $z$-direction~\cite{dixiao10,dixiao05} reads
\be\label{eq:densityBfield}
n(B,\mu)=\sum_{n\kv}\left(1+\frac{eB\Omega_{n\kv}}{\hbar}\right) f(E_{n\kv}-m_{n\kv}B-\mu)\,,
\ee 
where $\Omega_{n\kv} \hat\zv$ is the Berry curvature, $m_{n\kv}\hat\zv$ is the magnetic moment, and $E_{n\kv}$ is the energy of the Bloch state with band index $n$ and wave vector $\kv$.  The sum runs over all Bloch states, weighed with Fermi-Dirac distribution $f$, with chemical potential $\mu$.   Taking the derivative with respect to $B$ at constant $\mu$ and setting $B=0$ 
we obtain
\bea\label{eq:dndB}
e\left(\frac{\partial n}{\partial B}\right)_\mu &=& \frac{e^2}{\hbar}\sum_{n\kv}f(E_{n\kv}-\mu)\Omega_{n\kv}
\nonumber\\
&-&e\sum_{n\kv}f'(E_{n\kv}-\mu)m_{n\kv}, 
\eea
where $f'$ is the first derivative of $f$ with respect to energy.  

As a simple illustration of Eq.~(\ref{EinsteinH}), we focus on the intrinsic contribution to anomalous Hall responses where
the intrinsic anomalous Hall conductivity is the familiar~\cite{Ong,dixiao10}
\be\label{AHC}
\sigma_H =(e^2/\hbar)\sum_{n\kv}f(E_{n\kv}-\mu) \Omega_{n\kv}\,.
\ee
We note that this corresponds to the first term of Eq.~(\ref{eq:dndB}). Applying Eq.~(\ref{AHC}) and Eq.~(\ref{eq:dndB}) into Eq.~(\ref{EinsteinH}) we obtain the (Hall) diffusive response as
\be\label{DiffusiveAHE}
\bar\sigma_H \equiv \sigma_H-e\left(\frac{\partial n}{\partial B}\right)_\mu=-e\sum_{n\kv}f'(E_{n\kv}-\mu)m_{n\kv}\,.
\ee 
Strikingly, $\bar\sigma_H$ in Eq.~(\ref{DiffusiveAHE}) depends on the total magnetic moment around the Fermi surface. In contrast, the anomalous Hall conductivity $\sigma_H$ depends on the sum of the Berry curvature throughout the Fermi sea. This highlights the distinct origins of $\bar\sigma_H$ (diffusive: arising from Fermi surface $m_{n\vec k}$) and $\sigma_H$ (drift: arising from $\Omega_{n\vec k}$).
 
The anomalous Hall diffusion constant $D_H$ can be obtained from Eq.~(\ref{DiffusiveAHE}) in a straightforward fashion by noting $\partial n/\partial \mu = -\sum_{n\kv}f'(E_{n\kv}-\mu)$ (at $B=0$). This yields 
\be\label{HallDiffusionConstant}
eD_H = \Big[\sum_{n\kv}f'(E_{n\kv}-\mu)m_{n\kv}\Big]\Big/\Big[\sum_{n\kv}f'(E_{n\kv}-\mu)\Big]\,,
\ee
as the average magnetic moment on the Fermi surface for a metal. For non-degenerate semiconductors, where only a few carriers are present in the band, Eq.~(\ref{HallDiffusionConstant}) also applies -- it captures the average magnetic moment of thermally excited carriers. Interestingly, this can yield non-vanishing values of $D_H$ even at low temperatures. We note that extrinsic contributions to the Hall diffusion constant can be obtained in the same fashion as above by employing Eq.~(\ref{EinsteinH}).

We now proceed to assess the quantitative importance of our results. In so doing, we evaluate $\sigma_H$ and 
$\bar \sigma_H$ for a minimal two-band 
hamiltonian
\be
\label{eq:H}
H(\vec k) =  \varepsilon(\vec k) 1 + \vec d(\vec k) \cdot \boldsymbol{\sigma}, 
\ee
where $\vec k$ is a two-dimensional vector and $\boldsymbol{\sigma} = \sigma_x \hat{\vec{x}} + \sigma_y \hat{\vec{z}} + \sigma_z \hat{\vec{z}}$ are Pauli matrices describing a pseudo-spin degree of freedom. 
The energies of the two  bands are 
$
E_{\pm \vec k} = \varepsilon (\vec k) \pm |\vec d(\vec k)|,
$
and the Berry curvature and the magnetic moment are given by 
$
\Omega_{\pm \vec k} = \pm \epsilon_{ij} (d_z \partial_{i} d_x \partial_{j} d_y)/(2d^3)$ 
and 
$m_{\pm \vec k} = (e/\hbar) \epsilon_{ij}(d_z \partial_{i} d_x \partial_{j} d_y)/(2d^2) $ 
respectively~\cite{bernevig}, where $d=|\vec d(\vec k)|$, and $\epsilon_{ij}$ is the anti-symmetric tensor.

\begin{figure}
\includegraphics[width=\columnwidth]{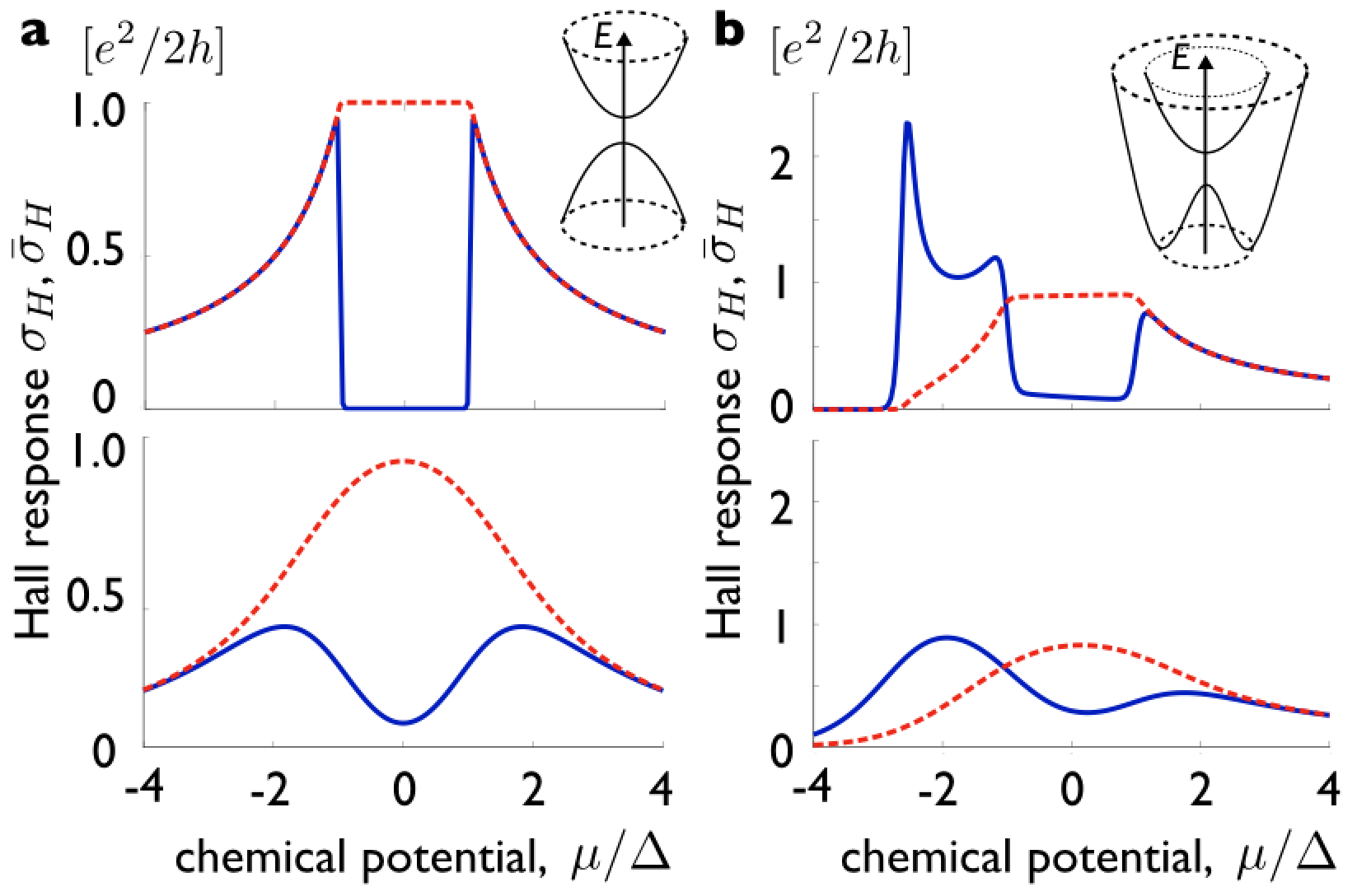}
\caption{Hall response for a mechanical force $\sigma_H$ (red dashed lines) and statistical force $\bar{\sigma}_H$ (blue solid lines) for various two-band models can exhibit contrasting behavior from Eq.~(\ref{AHC}) and Eq.~(\ref{DiffusiveAHE}) respectively, where we have focussed on the intrinsic contribution as an illustration. In panel ({\bf a}), the Hall responses for a gapped Dirac model with $\vec{d}(\vec k) = (v\hbar k_x, v \hbar k_y, \Delta)$ and $\varepsilon(\vec k) =0$ at various temperatures ($k_BT/\Delta = 0.01, 0.5$ for top and bottom) are displayed. In panel ({\bf b}), the Hall responses from a particle-hole asymmetric Eq.~(\ref{eq:H}) with $\vec{d}(\vec k) = (v\hbar k_x, v\hbar k_y, \Delta)$ and $\varepsilon(\vec k) =C |\vec k|^2$ is displayed at various temperatures ($k_BT/\Delta = 0.05, 0.5$ for top and bottom); we have used the dimensionless ratio $\Delta C/(v^2\hbar^2) =0.1$. Insets display sketch of the dispersion.}
\end{figure}

In Fig. 1, we plot the intrinsic $\sigma_H$ (dashed line) and $\bar\sigma_H$ (solid lines) using Eq.~(\ref{AHC}) and Eq.~(\ref{DiffusiveAHE}) for various $\vec d(\vec k)$ and $\varepsilon(\vec k)$ in Eq.~(\ref{eq:H}), various temperatures, and over a range of chemical potentials. For example, the broken time-reversal symmetry (TRS) Haldane model~\cite{Haldane} is captured by electrons around two gapped Dirac cones at different valleys. Since both valleys give the same Hall response, we concentrate on just one cone so that  $\vec d (\vec k) = (v\hbar k_x, v\hbar k_y, \Delta)$, and $\varepsilon(\vec k) =0$. Its Hall response is
plotted in panel (a).
The difference between $\bar\sigma_H$ and $\sigma_H$ is maximal for $\mu$ in the gap, where $\sigma_H=e^2/2h$ and $\bar \sigma_H=0$ at $T=0$. In the metallic state (e.g., $\mu$ crossing  the conduction band) the difference between $\bar\sigma_H$ and $\sigma_H$ vanishes at $T=0$~\cite{Pesin}. At finite temperature (bottom panel), differences between $\sigma_H$ and $\bar \sigma_H$ are no longer confined to chemical potential being in the gap, and appear for a range of chemical potentials close to the band edges. The same conclusion applies for the {\it valley} Hall conductivity~\cite{dixiao07,gorbachev14,tarucha15,zhang15,sid} in TRS preserving gapped Dirac systems e.g., gapped graphene on hexagonal Boron Nitride~\cite{dixiao07,gorbachev14}, transition metal dichacogenides~\cite{NingWang}. A similar analysis also applies for other valley/flavor Hall systems, e.g.,~\cite{sid}.

Next, we analyze the effect of breaking particle-hole symmetry by choosing $\varepsilon(\vec k) =C |\vec k|^2$ and $\vec{d}(\vec k) = (v\hbar k_x, v \hbar k_y, \Delta)$ in Fig. 1b; similar to that above, we focus on a single cone. Here differences between $\bar\sigma_H$ (solid) and $\sigma_H$ (dashed) can be significant in the metallic state even at low temperatures. 
Indeed, $\bar{\sigma}_H$ sharply rises at the band minimum $\mu \approx -2.6\Delta$ due to a large density of states close to a van Hove singularity (VHS), and then rises again just below the local maximum at $\mu=-\Delta$ and  the local minimum at $\mu=\Delta$ (see inset).  At $T=0$, $\bar{\sigma}_H$ close to the VHS diverges and has discontinuities at $\mu=\pm\Delta$; in the numerical plot of Fig.1b we have taken a low but finite temperature which cures these singularities (top panel). Strikingly, $\sigma_H$ close to the bottom of the lower band is small stemming from the small Berry flux sustained by the small Fermi sea. This presents an unusual situation wherein only small drift Hall currents are induced by an electric field (small $\sigma_H$), whereas large Hall diffusion currents can be easily driven by chemical potential gradients (large $\bar\sigma_H$) -- a clear sign of the Hall diffusion anomaly. 
Not only does $\bar{\sigma}_H$ sharply depart from $\sigma_H$, it also has a magnitude that surpasses $e^2/2h$ -- the maximum $\sigma_H$ (drift) Hall response that is permitted for the winding of $\vec d(\vec k)$ in this model. When $\mu$ approaches zero, the fortunes of $\sigma_H$ and $\bar{\sigma}_H$ are reversed, with the latter becoming small while the former maintains a large value close to $e^2/2h$. This vividly displays the rival origins and behavior of Hall drift and Hall diffusion.

{\bf Local vs global currents.} We emphasize that the anomalous diffusion constant $D_H$ is expected to work both in equilibrium and nonequilibrium situations, and will manifest in conventional transport experiments where the current is measured at global contacts of a macroscopic device. Such {\it global} currents, $I$, (also known as transport currents) can be obtained as the flux of $\delta \vec j (\vec r)$ through an appropriate cross section of the device.  

Significant differences between local current density and global currents emerge when one considers their behavior in equilibrium and non-equilibrium situations. In both
equilibrium and out-of-equilibrium settings, a finite local current density $\delta \vec j (\vec r)$ can exist. Nevertheless, when the system is at equilibrium, {\it global} currents collected at macroscopic contacts must necessarily vanish in the thermodynamic size limit. 
This follows from the fact that the existence of a finite current flux at equilibrium would allow the system to reduce its free energy by shifting the phase of the wave function along the direction of the current.  
Thermodynamic stability therefore requires the current flux in equilibrium to obey the inequality~\cite{Bohm1949,Thouless1993} (see {\bf SI} for a full derivation for the convenience of the reader~\cite{SI}),
\be
|I|_{\rm eq}\leq \frac{e h}{2 m_{\rm eff} L^2}, 
\label{eq:Bohm}
\ee
where $L$ is the (macroscopic) length of the circuit, and $m_{\rm eff}$ is the effective mass of the electrons.
Notice that the statement leaves open the possibility of mesoscopic equilibrium currents, which will vanish only in the thermodynamic limit $L\to\infty$. 

The physical picture is the following: as an electronic system relaxes to thermodynamic equilibrium, the total electric and chemical potential adjust to a (global) equilibrium configuration with profiles $\varphi (\vec r), \mu^{\rm eq}(\vec r)$ across the entire device so that $|I|_{\rm eq}$ vanishes, Eq.~(\ref{eq:Bohm}). As a result of this adjustment, even when finite local current density persists at equilibrium they automatically integrate to zero when one calculates the flux of the current density through a cross section of the system. 
In contrast, nonequilibrium currents are not constrained by Eq.~(\ref{eq:Bohm}) and will sum to give a non-vanishing net current flux through a device cross section. This is because a nonequilibrium (global) configuration of electric and chemical potentials do not yield the lowest free energy.

Crucially, Eq.~(\ref{eq:Bohm}) tells us that the distinction between equilibrium 
and nonequilibrium currents 
arises {\it globally} and cannot be deciphered locally. For e.g., consider an electron gas pushed out-of-equilibrium where applied $\delta \mu(\vec r), \delta \varphi (\vec r)$ profiles do not obey Eq.~(\ref{EquilibriumCondition}).
The transverse local charge current density induced reads 
\be
\delta \vec j_H (\vec r) = \sigma_H (\vec r) \boldsymbol{\nabla} \times e\delta \varphi (\vec r) \hat{\vec{z}} -  \bar{\sigma}_H(\vec r) \boldsymbol{\nabla} \times \delta \mu (\vec r) \hat{\vec{z}},   
\label{eq:TotalHall}
\ee
where we have substituted Eq.~(\ref{EinsteinH}) into Eq.~(\ref{eq:fullcurrent}), and cycled the vector product; $H$ denotes transverse current and we have explicitly displayed the $\vec r$ dependence of all quantities for clarity. 
In such a setting, it is no longer meaningful, or even possible, to delineate between equilibrium and nonequilibrium components of the local current density. This is because we only know a local current density is part of a global equilibrium pattern when the entire pattern sums to zero flux; pairs of currents that cancel can occur far apart from each other. Indeed, a direct measurement of the current density, which is possible via accurate measurements of the magnetic field (see e.g., Ref.~\cite{Uri2020}), would always give the {\it total} current density at a point $\rv$, regardless of whether the net current flux is zero or not.

It was noted in Ref.~\cite{Cooper1997,dixiao06} that a transverse current density that is 
completely expressible as the curl of a vector field [e.g., $\boldsymbol{\nabla} \times \vec M (\vec r)$] 
would necessarily have zero flux through a cross section of the system for a vanishing $\vec M (\vec r) = 0$ outside the system. 
Importantly, none of the terms in Eq.~(\ref{eq:TotalHall}) are locally expressible as a curl and 
do not have to vanish when integrated through a cross section of the system. Crucially, both drift and diffusion terms in Eq.~(\ref{eq:TotalHall}) appear on the same footing. As a result, determining the net Hall current (measured at global contacts) requires integrating the complete $\delta \vec j_H (\vec r)$ over an {\it entire} cross section. 

Depending on the profile of applied $\delta \varphi(\vec r)$ and $\delta \mu(\vec r)$ across the device cross-section, the
net Hall current drawn will be determined by $\sigma_H$, $\bar{\sigma}_H$, or both. As an illustration, consider a uniform Hall bar with a homogeneous $\sigma_H(\vec r) = \sigma_H^{(0)}$, $\bar{\sigma}_H(\vec r) = \bar{\sigma}_H^{(0)}$ for $\vec r$ inside the device, but vanishes for $\vec r$ outside the device. When $\nabla \delta \varphi (\vec r)$ is directly sustained by applying an external electric field (but keeping $\nabla \delta \mu (\vec r) =0$) across a Hall bar device, we obtain a net bulk mechanical drift Hall current $I_H = \sigma_H^{(0)} \Delta\varphi$, where $\Delta \varphi$ is the electric potential drop across the cross section. Similarly, $\nabla \delta \mu (\vec r)$ can be induced (keeping $\nabla \delta \varphi (\vec r) = 0$) by photoexcitation using a local laser spot. This can create a local carrier density gradient that drives photo-induced diffusion currents as in Eq.~(\ref{DiffusionCurrent1}), yielding a net bulk Hall diffusion current $I_H = - \bar{\sigma}_H^{(0)} \Delta\mu$, where $\Delta \mu$ is the chemical potential imbalance across the cross section; this can be collected as a chiral photocurrent~\cite{ShockleyRamo}. 

Aside from its significant influence on global charge transport, described above, we note that there are other means with which to extract out Hall diffusion and the Hall diffusion anomaly. For example, by directly measuring the current density accompanying minority-carrier diffusion in a Haynes-Shockley-type experiment~\cite{Haynes-Shockley1949}, or by tracking the  current density profile using sensitive spatially resolved magnetometry~\cite{Uri2020}. 

On a fundamental level, we have pointed out the existence of two different transport coefficients governing the linear responses of electronic systems to mechanical and statistical forces. This manifests as a Hall diffusion anomaly (arising under very general macroscopic considerations) and exhibit distinct Hall drift $\sigma_H$ and Hall diffusion $\bar{\sigma}_H$ coefficients that can differ from each other by significant amounts. 
We expect the anomaly to become even more important when considering the Hall effect for non-charged quasiparticles, such as excitons~\cite{Iwasa} or anomalous thermal transport~\cite{Shastry}. In all these cases there are essentially no mechanical forces and the transport is driven in its entirety by statistical forces: the gradient of the chemical potential in the first two cases, the gradient of the temperature in the last.

\vspace{2mm}
{\bf Acknowledgements} - We gratefully acknowledge useful conversations with Mark Rudner. J.C.W.S. acknowledges support from Singapore National Research Foundation (NRF) under NRF fellowship award NRF-NRFF2016-05, 
and a Singapore MOE Academic Research Fund Tier 3 Grant MOE2018-T3-1-002, 
as well as the Aspen Center for Physics, 
which is supported by National Science Foundation grant PHY-1607611, where part of this work was performed. G.V. acknowledges support from the CA2DM, where this work was initiated, and from DOE Grant DE-FG02-05ER46203.


\clearpage
\newpage

\renewcommand{\theequation}{S-\arabic{equation}}
\renewcommand{\thefigure}{S-\arabic{figure}}
\renewcommand{\thetable}{S-\Roman{table}}
\makeatletter
\renewcommand\@biblabel[1]{S#1.}
\setcounter{equation}{0}
\setcounter{figure}{0}

\twocolumngrid




\section{Supplementary Information for ``Hall diffusion anomaly and transverse Einstein relation''} 

\section{Physical reason for vanishing longitudinal current density at equilibrium}

In the main text, we derived the Einstein relations for both longitudinal and transverse components of current. In this section, we give a simple physical reason why the longitudinal Einstein relation has to hold. 

Like any vector field, the current density can be expressed as the sum of a longitudinal and a transverse component: the longitudinal component has zero curl, while the transverse component has zero divergence. Only the longitudinal component of the current density is connected to the time derivative of the density by the continuity equation.  At equilibrium, the time derivative of the density vanishes, implying that the longitudinal component of the current must also vanish. As a result, Eq.~(\ref{EinsteinRelationLongitudinal}) of the main text must be satisfied. However, as we discuss in the main text, the chain of reasoning leading to Eq.~(\ref{EinsteinRelationLongitudinal}) fails for the transverse components of response tensors. 

\section{Generalization to three-dimensions}

In the main text, we derived the Einstein relations (of the longitudinal and transverse components) for current in a two-dimensional system. Such relations also hold more generally for three dimensions, $(x,y,z)$. As we describe below, this can be done in much the same way as in the main text. For simplicity of notation we specify the magnetization to be pointing in the $\hat{\vec{z}}$ direction. As such, Hall currents (for both drift and diffusive types) do not flow along $\hat{\vec{z}}$; instead they flow only in the $x$-$y$ plane. 

Similar to that described in the main text, the drift current can be characterized as 
\be\label{eq:S-Drift}
\widehat{\boldsymbol{\mathcal{L}}} \big[-\nablabold\delta\varphi(\rv)\big]  = -\left(\begin{array}{ccc} \sigma_{xx} & -\sigma_H & 0 \\ \sigma_H & \sigma_{yy} & 0 \\ 0 & 0 & \sigma_{zz}\end{array}\right)\left(\begin{array}{ccc} \partial_x \delta\varphi\\ \partial_y \delta\varphi\\ \partial_z \delta\varphi\end{array}\right),
\ee
with longitudinal conductivity $\sigma_{xx,yy,zz}$ along the $x,y,z$ directions, and  transverse (Hall) conductivity $\sigma_H$. In a similar fashion, we express the tensor $\widehat{\boldsymbol{\mathcal{N}}}$ as %
\be
\label{eq:S-Diff}
\widehat{\boldsymbol{\mathcal{N}}}\big[\nablabold \delta \mu(\vec r) \big] =e\frac{\partial n}{\partial\mu} \left(\begin{array}{ccc} D_{xx} & -D_H & 0 \\ D_H & D_{yy} & 0 \\ 0 & 0 & D_{zz}\end{array}\right)\left(\begin{array}{ccc} \partial_x \delta\mu \\ \partial_y \delta\mu \\ \partial_z \delta\mu \end{array}\right),  
\ee
where $D_{xx,yy,zz}$ are the diffusion constants along the $x,y,z$ directions respectively. 

We can follow the same reasoning as the main text. First, we recall that in the linear response regime, Eq.~(\ref{eq:S-Drift}) and (\ref{eq:S-Diff}) are valid with {\it constant} and same transport coefficients ($\sigma$ and $D$) in both equilibrium and out-of-equilibrium settings. Next we focus on the situation when the system is allowed to relax to equilibrium, wherein Eq.~(\ref{EquilibriumCondition}) of the main text applies. Indeed, the only equilibrium current density that is allowed to flow is then given by Eq.~(\ref{deltaj-equilibrium}) of the main text. 

As in the main text, equating the sum of the drift Eq.~(\ref{eq:S-Drift}) and diffusion (\ref{eq:S-Diff}) current density to the equilibrium current density sustained (when the $\delta \mu(\vec r)$ and $\delta \varphi(\vec r)$ satisfy Eq.~(\ref{EquilibriumCondition}) of the main text yields the Einstein relations 
\be
\left(\begin{array}{ccc} \sigma_{xx} \\ \sigma_{yy} \\ \sigma_{zz} \end{array}\right)= e^2\frac{\partial n}{\partial\mu} \left(\begin{array}{ccc} D_{xx} \\ D_{yy} \\ D_{zz} \end{array}\right)
\ee
and the transverse Einstein relation 
\be
\sigma_H-e\left(\frac{\partial n}{\partial B_z}\right)_\mu = e^2\frac{\partial n}{\partial\mu} D_H \,.
\ee
where $B_z$ is magnetic field in the $\hat{\vec{z}}$ direction, i.e. along the same direction as the magnetization. 

\section{Bloch theorem and vanishing of net equilibrium current in the thermodynamic limit}

In this section we show that the net current traversing any cross section of an electronic system at thermal equilibrium vanishes in the thermodynamic limit.  This conclusion applies as well to the net current that flows through any additional wires which one may attach to the primary system in an attempt to extract current  from it.   We thus conclude that any  currents which may locally circulate in a system at equilibrium cannot be extracted, i.e., cannot generate a net flow of current on a macroscopic scale.  Notice that this statement leaves  open the possibility of mesoscopic equilibrium currents, such as persistent currents in mesoscopic rings~\cite{Levy,Chandrasekhar}.

\begin{figure}[b]
\label{FigS2}
\begin{center}
\includegraphics[width=\columnwidth]{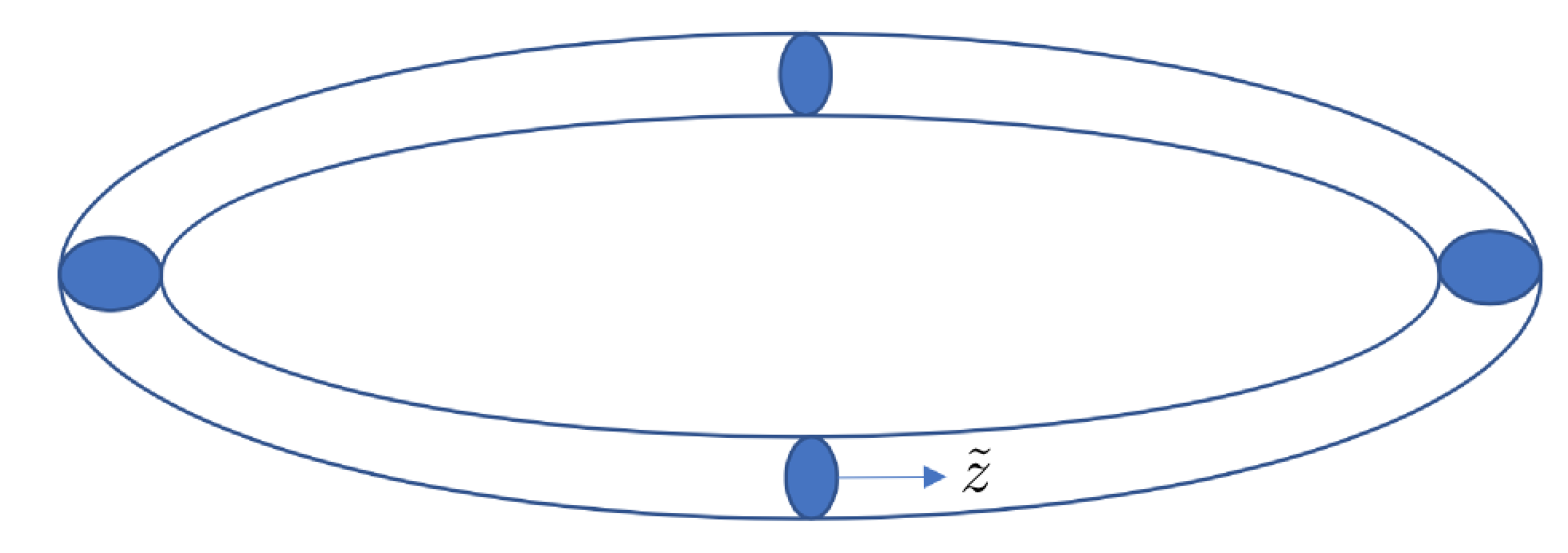}
\caption{Schematic of a cylindrical wire used in the proof of the ``Bloch'' theorem.  The modular coordinate $\tilde{z}$ runs around the loop, with $0<\tilde{z}<L$.}
\end{center}
\end{figure}

Our argument closely parallels the theorem proved by Bohm~\cite{Bohm1949}, following an earlier suggestion by Bloch, for systems in the ground state (see also Ref.~\cite{Thouless1993}).  However, we explicitly extend the proof to finite temperature.  

Consider a 3D wire, such as shown in Fig.~\ref{FigS2}, in thermal equilibrium with density matrix $\boldsymbol{\rho}$.  The current density satisfies the condition $\nabla \cdot \jv(\rv)=0$.  This condition does not mean that $\jv=0$ everywhere.  It is certainly compatible with the existence of local circulating currents, and furthermore it is compatible with the presence of a net current $I$ flowing through any cross-section of the system -- the same for any cross section.   Without loss of generality, we can assume that the system is a cylinder of constant cross section.  Any deviation from this ideal geometry can be corrected by a smooth deformation, which does not change the net current flowing in the loop.  Let us introduce cylindrical coordinates $\rv =(\tilde{z},\tilde{\rho},\phi)$ where the modular coordinate $\tilde{z}$  ($0<\tilde{z}<L$)  keeps track of position along the wire.  It immediately follows from the constancy of the flux through cross sections at different $\tilde{z}$ that the volume integral of the current density satisfies the condition
\be
\int j_{\tilde{z}}(\rv) dV=IL
\ee
where the integral runs over the volume of the wire and $j_{\tilde{z}}$ is the $\tilde{z}$-component of the current density. 
Now let us multiply all the wave functions of the system (ground state and excited states) by a phase factor $e^{\pm 2\pi i\sum_i \tilde{z}_i/L}$ where the sum runs over all the electrons.  Notice that the symmetry and single-valuedness of the wavefunctions are preserved in this transformation. If the original wave functions were eigenfunctions of the hamiltonian
\be\label{Hamiltonian}
H=\sum_i\left\{\frac{p_i^2}{2m_{\rm eff}} +V(\rv_i)\right\} +\sum_{i <j}\frac{e^2}{|\rv_i-\rv_j|}\,,
\ee
where $V(\rv)$ is a generic potential, and $m_{\rm eff}$ is the effective mass of the electrons, then the transformed wave functions are eigenfunctions of the hamiltonian 
\ber H' &=& e^{\pm 2\pi i\sum_i \tilde{z}_i/L}He^{\mp 2\pi i\sum_i\tilde{z}_i/L}\nn\\
&=& H\pm\frac{h}{eL} \int  j_{\tilde{z}}(\rv) dV +\frac{h^2N}{2m_{\rm eff}L^2}\,,
\eer
with the same eigenvalues ($N$ is the number of electrons).
Now we use the inequality~\cite{Feynman}
\be
F'\leq F+\langle H'-H \rangle
\ee
where  $F'$ and $F$ are the equilibrium free energies associated with the hamiltonians $H'$ and $H$ respectively and the thermal average of an observable $A$ (in this case $H'-H$) is defined as $\langle A \rangle \equiv Tr[\boldsymbol{\rho} A]$ , i.e., the average in the canonical ensemble of the hamiltonian $H$.  Since $H'$ and $H$ differ by a unitary transformation, it must be the case that $F'=F$ (the trace of $e^{-\beta H}$ is the same as the trace of $e^{-\beta H'}$).  Using the explicit form of the difference $H'-H$ we obtain the inequality
\be
\pm \frac{hI}{e}+N \frac{h^2}{2m_{\rm eff}L^2}\geq 0   \rightarrow   |I|\leq \frac{e h}{2 m_{\rm eff}L^2}\,.
\ee
The right hand side tends to zero in the limit $L\to \infty$.  This proves that the integrated current $I$ vanishes in equilibrium in the thermodynamic size limit. 

The argument can be generalized to multiply connected circuits  consisting of several loops.  These circuits consist of a primary loop, which closes on itself as before, and parallel branches that connect to the primary loop at two points.  We extend our coordinate system to the parallel branches of the circuit.   We then multiply the wave functions by a phase factor that grows linearly with $\tilde{z}$ in each branch.  This is allowed if the phase changes of the wave function along the parallel branches  are identical to the phase changes in the primary branch between the same end points, or differ by a multiple of $2\pi$. The variational principle, applied to the situation in which the phase change is applied to only one of the parallel branches, leads to the conclusion that the net current flux is zero in that branch, as well as in the primary loop, in the thermodynamic limit.

\end{document}